\documentclass[aps,twocolumn]{revtex4}
\usepackage{epsfig}
\usepackage{graphicx,color,dcolumn}
\usepackage{epstopdf}
\usepackage{amsmath,amssymb}
\usepackage{multirow}
\usepackage{diagbox}
\usepackage{changes}
\usepackage{booktabs}
\usepackage{threeparttable}
\usepackage{subfigure}
\usepackage{hyperref}

\begin{document}
\title{Exploring the nature of $\eta_{1}(1855)$ and it's partner in a chiral quark model}
\author{
Yue Tan\textsuperscript{a},
Yu-Heng Wu\textsuperscript{a},
Qi Huang\textsuperscript{b},
Xiaoyun Chen\textsuperscript{c},
Xiaohuang Hu\textsuperscript{d},
Youchang Yang\textsuperscript{e},
Jialun Ping\textsuperscript{b}
}
\email[E-mail: ]{huangqi@nnu.edu.cn (Corresponding author) }
\email[E-mail: ]{jlping@njnu.edu.cn (Corresponding author)}
\affiliation{\textsuperscript{a}Department of Physics, Yancheng Institute of Technology, Yancheng 224000, People's Republic of China }
\affiliation{\textsuperscript{b}Department of Physics, Nanjing Normal University, Nanjing 210023, People's Republic of China }
\affiliation{\textsuperscript{c}College of Science, Jinling Institute of Technology, Nanjing 211169, People's Republic of China }
\affiliation{\textsuperscript{d}Department of Physics, Changzhou Vocational Institute of Engineering, Changzhou 213164, People's Republic of China }
\affiliation{\textsuperscript{e}Department of Physics, Guizhou University of Engineering Science, Bijie 551700,  People's Republic of China.}

\date{\today}


\begin{abstract}
Inspired by the experimental discoveries of \(X(3872)\) (\(c\bar{q}\)-\(q\bar{c}\)) and \(T_{cc}\) (\(c\bar{q}\)-\(c\bar{q}\)), we do a systematically study on their orbit-excited strange partners, i.e., the \(K \bar{K}_1\) (\(q\bar{s}\)-\(s\bar{q}\)) system, which may be related with the recently observed \(\eta_1(1855)\), and the \(K K_1\) (\(q\bar{s}\)-\(q\bar{s}\)) system. Within the framework of an accurate few-body calculation method (GEM), we employ the chiral quark model to simultaneously consider the molecular and diquark structures of these two multiquark systems and include their channel coupling effects. Our results show that the \(K \bar{K}_1\) system remains a scattering state. On the other hand, due to the presence of a good-diquark structure in the \(K K_1\) system, we obtain a bound state in the coupled-channel calculation, where the\(\pi\)-meson and \(\sigma\)-meson. The inter-quark distance indicates that it is a compact four-quark structure.
\end{abstract}

\maketitle

\section{Introduction}\label{introduction}

To give proper explanation on the newly observed exotic states in experiments, decoding the inner hadronic structures of them is one of the most intriguing and important topics in hadron physics. In the quark model, exotic states can be classified into two categories. One is that their properties cannot be adequately described by the traditional quark model. For example, $X(3872)$ \cite{Godfrey:1985xj,Tan:2019qwe} is interpreted as a $c\bar{c}$ state with $1^{++}$ in the traditional quark model, yet its mass exceeds the experimentally observed value by several tens of MeV. The other is that their quantum numbers cannot be achieved as a meson or baryon, such as the $Z_c(3900)$ \cite{Zhang:2013aoa,Chen:2013coa,Guo:2013xga,Liu:2013vfa} with $1^{+-}$. Studying these exotic hadrons can deepen our understanding of hadron spectroscopy, and, in turn, lead to a better comprehension on QCD.

Recently, the BESIII collaboration observed a new structure $\eta_1(1855)$ in the $J/\psi \rightarrow \gamma \eta \eta^{\prime}$ process with a significance of 19$\sigma$ \cite{BESIII:2022riz,BESIII:2022iwi}. Its mass and width are measured to be $1855 \pm 9^{+6}_{-1}$ MeV and $188 \pm18 ^{+3}_{-8}$ MeV, respectively. Due to its unusual quantum numbers, it cannot be described by a conventional quark-antiquark structure. Moreover, since its mass is very close to the threshold for $K\bar{K}_{1}(1400)$, a possible assignment is that it may be a molecular state, specifically a $K \bar{K}_{1}(1400)$ state.

Currently, theoretical studies of $\eta_1(1855)$ can be broadly classified into two categories, one interprets it as a hybrid state such as $q\bar{q}g$ \cite{Narison:1999hg,Shastry:2022mhk,Chen:2023ukh,Iddir:1988jd,Benhamida:2019nfx,Eshraim:2020ucw,Zhang:2025xee,Dudek:2010wm,Qiu:2022ktc,Chen:2022qpd,Narison:2009vj,Ma:2025cew}, while the other proposes it as a molecular $K \bar{K}_{1}$ ($q\bar{s}$-$s\bar{q}$) state \cite{Chen:2008qw,Wan:2022xkx,Yang:2022rck,Dong:2022cuw,Yan:2023vbh}. For example, the authors in \cite{Shastry:2022mhk} studied the decays of the $J^{PC} = 1^{-+}$ hybrid nonet using a Lagrangian invariant under flavor symmetry and found that the light isoscalar must be significantly narrow, while the width of the heavy isoscalar can match the recently observed $\eta_1(1855)$. Through the study of the corresponding decay modes of $\eta_1(1855)$, the authors in \cite{Chen:2022qpd} explained it as an $s\bar{s}g$ hybrid meson and found that the QCD axial anomaly enhances the decay width of the $\eta\eta'$ channel, although this mode is strongly suppressed by the small P-wave phase space. This conclusion is confirmed in \cite{Qiu:2022ktc}, which is based on the flux tube model.  Recently, Ma et al. \cite{Ma:2025cew} included effective gluons in the constituent quark model and recalculated the \(1^{-+}\) light hybrids, including \(q\bar{q}g\), \(s\bar{s}g\), and \(q\bar{s}g\). Their results provide a good explanation of the energy spectrum for \(\pi(1600)\) and \(\eta_1(1855)\), but the candidate for \(\eta_1(1855)\), mainly \(s\bar{s}g\), has a nearly zero decay width to \(\eta \eta'\), which can not explain the experimental observation of \(\eta_1(1855)\) on $J/\psi \rightarrow \gamma \eta \eta^{\prime}$ process. On the other hand, some works suggest that $\eta_1(1855)$ could be a molecular state of $K\bar{K}_1$. For instance, in Ref.~\cite{Yang:2022rck}, the authors studied the radiative and strong decays of the S-wave $K\bar{K}_1(1400)$ molecular state within the effective Lagrangian approach and found that its decay width is consistent with the experimentally observed $\eta_1(1855)$, while the study in Ref.~\cite{Dong:2022cuw} obtained the bound state of $K\bar{K}_1(1400)$ based on the one-boson exchange model. Similarly, Ref.~\cite{Yan:2023vbh} conducted a similar study by using the Bethe-Salpeter equation. In Ref.~\cite{Wan:2022xkx}, the tetraquark interpretation for the structure of $\eta_1(1855)$ is examined, where the observed $\eta_1(1855)$ could be embedded in the $[1_c] \bar{s}s \otimes [1_c] \bar{q}q$ configuration.  However, within the framework of QCD sum rules, a recent study by the authors of Ref.\cite{Liu:2024lph} conducted detailed calculations of the correlation functions for the $K\bar{K}_1$ system, and their find no evidence for a bound state formation. 

Apparently, the tetraquark explanation on $\eta_1(1855)$ indicates that it can be seen as a orbital-excited strange partner of \(X(3872)\). Since the observation of \(X(3872)\), with its quark configuration as \(c\bar{q}\)-\(q\bar{c}\), has further inspired experimental physicists to discover a similar multiquark state, \(T_{cc}\), with a quark composition of \(c\bar{q}\)-\(c\bar{q}\). Therefore, in this paper, we investigate not only the S-wave \(K \bar{K}_1\) molecular state (\(q\bar{s}\)-\(s\bar{q}\)), but also the S-wave \(K K_1\) molecular state (\(q\bar{s}\)-\(q\bar{s}\)). Considering that both the \(q\bar{s}\)-\(s\bar{q}\) and \(q\bar{s}\)-\(q\bar{s}\) tetraquark systems are composed of light and strange quarks, we propose a framework based on SU(3) symmetry within the chiral quark model (ChQM). To calculate the properties of these tetraquark systems, we employ the Rayleigh-Ritz variational method combined with the Gaussian expansion method (GEM), which allows us to expand each relative motion in the system in terms of Gaussian basis functions. In addition, we take into account the mixing effects between the molecular and diquark-antidiquark configurations.

The paper is organized as follows, after introduction, section \ref{wavefunction and chiral quark model} presents the details of our model, section \ref{result} presents the numerical results, and the final section \ref{Summary} is devoted to the summary.

\section{Model setup} \label{wavefunction and chiral quark model}
\subsection{Chiral quark model}
In the chiral quark model \cite{Vijande:2004he,Yang:2009zzp,Tan:2020ldi}, the potential part of the Hamiltonian consists of the color confinement potential, one-gluon-exchange potential, Goldstone boson-exchange potential, and the \(\sigma\)-meson exchange. Since we are considering excited meson systems as \(K\bar{K}_1\) (\(KK_1\)), the Hamiltonian also includes spin-orbit coupling. As all, the Hamiltonian of the ChQM used here is given as
\begin{eqnarray}
\nonumber
H &=&\sum_{i=1}^4 ( m_i +\frac{\vec{p}_{i}^{~2}}{2 m_{i}} ) -T_{CM} +  \sum_{i<j=1}^4( V_{con}(r_{ij})  \\
&&+V_{oge}(r_{ij})+V_{\chi}(r_{ij}) +V_{\sigma }(r_{ij})) ,
\end{eqnarray}
where \(\chi = \pi, K, \eta\), and \(T_{CM}\) is the kinetic energy operator for the center-of-mass motion of the whole system. Considering that the \(K\bar{K}_1\) system is composed of light quarks, the Goldstone boson-exchange potential and \(\sigma\)-meson exchange potential, which reflect the spontaneous breaking of chiral symmetry, play a significant role, and they can be explicitly written as
\begin{eqnarray}
\nonumber
V_{\chi}  &=& v_{\pi}({r_{ij}}) \sum_{a=1}^{3} \lambda_i^a \lambda_j^a + v_{K}({{r}_{ij}}) \sum_{a=4}^{7} \lambda_i^a \lambda_j^a \\
  &&+ v_{\eta}({{ r}_{ij}}) \left[ \cos\theta_{P} (\lambda_i^8 \lambda_j^8) - \sin\theta_{P} \right], \\
\nonumber
v_{\chi=\pi,K,\eta}(r_{ij}) & = &\frac{g^2_{ch}}{4\pi} \frac{m_{\chi}^2}{12 m_i m_j} \frac{\Lambda^2_{\chi}}{\Lambda^2_{\chi} - m_{\chi}^2} m_{\chi} [ Y(m_{\chi} r_{ij}) \nonumber\\
&&- \frac{\Lambda_{\chi}^3}{m_{\chi}^3} Y(\Lambda_{\chi} r_{ij}) ] (\vec{\sigma}_i \cdot \vec{\sigma}_j).
\end{eqnarray}
For $\sigma$ exchange, the potential consists of central part and spin-obrit part as
\begin{eqnarray}
V_{\sigma}^C  &=& -\frac{g^2_{ch}}{4\pi} \frac{\Lambda_s^2 m_s}{\Lambda_s^2 - m_s^2} \left[ Y(m_s r_{ij}) - \frac{\Lambda_s}{m_s} Y(\Lambda_s r_{ij}) \right], \\
V_{\sigma}^{SO}  &=& -\frac{g^2_{ch}}{4\pi} \frac{\Lambda_s^2}{\Lambda_s^2 - m_s^2} \frac{m_s^3}{2 m_i m_j} [ G(m_s r_{ij}) - \frac{\Lambda_s^3}{m_s^3} \\
\nonumber&&\times G(\Lambda_s r_{ij}) ]\vec{L} \cdot \vec{S},
\end{eqnarray}
where $\boldsymbol{\lambda}^{a}$ are the \(SU(3)\) flavor Gell-Mann matrices, $m_{\chi= \pi, K, \eta}$ is the mass of the Goldstone bosons, and \(\Lambda_{\chi= \pi, K, \eta}\) are the cut-offs. The \(Y(x)\) is the standard Yukawa function, defined as \(Y(x) = e^{-x}/x\), and \(G(x)\) is given by \(G(x) = (1 + 1/x) Y(x)/x\). Finally, the chiral coupling constant \( g_{ch} \) is determined from the \( \pi N N \) coupling constant via the relation
\begin{eqnarray}
\frac{g^2_{ch}}{4\pi} = \left( \frac{3}{5} \right)^2 \frac{g^2_{\pi NN}}{4\pi} \frac{m^2_{u,d}}{m^2_N}.
\end{eqnarray}

In the quark model, there are usually three commonly used forms of confinement potential, i.e., linear confinement\cite{Yang:2011rp}, quadratic confinement\cite{Tan:2020ldi}, and color-screened confinement \cite{Vijande:2004he}. These are all effective in describing the energy of the ground state. However, in our study, which involves the excited states of the $K$ meson, particularly the $K \bar{K}_1$ ($K K_1$) tetraquark state, the color-screened confinement potential is more suitable. Its form is given as follows
\begin{eqnarray}
    V_{\text{con}}^{C} &=  ( -a_c(1 - e^{-\mu_c r_{ij}}) + \Delta ) \boldsymbol{\lambda}_i^c \cdot \boldsymbol{\lambda}_j^c, \\
    \nonumber V_{\text{con}}^{SO} &= -\boldsymbol{\lambda}_i^c \cdot \boldsymbol{\lambda}_j^c \frac{a_c \mu_c e^{-\mu_c r_{ij}}}{4m_i^2 m_j^2 r_{ij}} [ ((m_i^2 + m_j^2)(1 - 2a_s)  \\
   \nonumber & + 4m_i m_j (1 - a_s)) (\vec{S}_{+} \cdot \vec{L}) + ((m_j^2 - m_i^2)\\
    &(1 - 2a_s)) (\vec{S}_{-} \cdot \vec{L}) ].
\end{eqnarray}
where $\boldsymbol{\lambda}^{c}$ are $SU(3)$ color Gell-Mann matrices, $a_{c}$, $\mu_c$  are model parameters. For the one-gluon-exchange, it provides both intermediate and long-range attraction for the multiquark systems here, and its form is given as
\begin{eqnarray}
    V_{\text{oge}}^{C} &=& \frac{\alpha_s}{4} \boldsymbol{\lambda}_i^c \cdot \boldsymbol{\lambda}_j^c
    \left[\frac{1}{r_{ij}} - \frac{1}{6m_i m_j} \boldsymbol{\sigma}_i \cdot \boldsymbol{\sigma}_j \frac{e^{-r_{ij}/r_0(\mu_{ij})}}{ r_{ij} r_0^2(\mu_{ij})} \right], \\
  \nonumber   V_{\text{oge}}^{SO} &=& - \frac{1}{16} \frac{\alpha_s}{m_i^2 m_j^2} \boldsymbol{\lambda}_i^c \cdot \boldsymbol{\lambda}_j^c
    [ \frac{1}{r_{ij}^3} - \frac{e^{-r_{ij}/r_g(\mu)}}{r_{ij}^3}  \\
  \nonumber   &&\times( 1+ \frac{r_{ij}}{r_g(\mu)} ) ]  [ ((m_i + m_j)^2 + 2 m_i m_j) (\vec{L} \cdot \vec{S}) \\
    &&+ (m_j^2 - m_i^2)(\vec{L} \cdot \vec{S}_-)].
\end{eqnarray}
where \(\alpha_s\) is the QCD-inspired strong coupling constant, which is determined by fitting experimental meson data. All of model parameters are shown in Table~\ref{modelparameters}, and the meson masses are shown in Table~\ref{modelmass}.

\begin{table}[t]
\begin{center}
\caption{Quark model parameters (\(m_{\pi}=0.7\) \(fm^{-1}\), \(m_{\sigma}=3.42\) \(fm^{-1}\), \(m_{\eta}=2.77\) \(fm^{-1}\), \(m_{K}=2.51\) \(fm^{-1}\)).\label{modelparameters}}
\begin{tabular}{cccc}
\hline\hline\noalign{\smallskip}
Quark masses   &$m_u=m_d$(MeV)     &313  \\
               &$m_{s}$(MeV)         &555  \\
Goldstone bosons   
                   &$\Lambda_{\pi}=\Lambda_{\sigma}(fm^{-1})$     &4.2  \\
                   &$\Lambda_{\eta}=\Lambda_{K}(fm^{-1})$     &5.2  \\
                   &$g_{ch}^2/(4\pi)$                &0.54  \\
                   &$\theta_p(^\circ)$                &-15 \\
Confinement             &$a_{c}$ (MeV)     &430 \\
                        &$a_{s}$ (MeV)     &0.777 \\
                        &$\mu_{c}$($fm^{-1}$)     &0.7 \\
                        &$\Delta$(MeV)       &181.1 \\
OGE                 & $\alpha_{qq}$        &0.54 \\
                    & $\alpha_{qs}$        &0.48 \\
                    & $\alpha_{ss}$        &0.42 \\
                    &$\hat{r}_0$(MeV)    &28.17 \\
                    &$\hat{r}_g$(MeV)    &34.5 \\
\hline\hline
\end{tabular}
\end{center}
\end{table}

\begin{table}[t]
\begin{center}
\caption{Mass of $K$-meson family.\label{modelmass}}
\begin{tabular}{cccccc}
\hline\hline\noalign{\smallskip}
~~~$J^{P}$~~~  & ~~~ Spin   ~~~&~~~ state           ~~~& ~~~Energy  ~~~& ~~~ Mixed ~~~   \\
$0^{-}$  &  0      &  $K(495)     $  &  485                \\
$1^{-}$  &  0      &  $K^{*}(892) $  &  908                \\
$0^{+}$  &  1      &  $K_{0}(800) $  &  1320               \\
$1^{+}$  &  1      &  $K_{1}(1400)$  &  1410  & 1517       \\
$1^{+}$  &  0      &  $K_{1}(1271)$  &  1393  & 1267       \\
$2^{+}$  &  1      &  $K_{2}(1430)$  &  1414               \\
\hline\hline
\end{tabular}
\end{center}
\end{table}

\subsection{The wave function of $K\bar{K}_1$ and $KK_1$ system}

In the quark model, the total tetraquark wave function \(\Psi(r_{ij})^{i,j,k,l}\) is constructed by a direct product of the spatial wave function \(\phi_L\), the spin wave function \(\sigma_s\), the flavor wave function \(\zeta_{I}\), and the color wave function \(\chi_{c}\), and then multiplied by the antisymmetrization operator \(\mathcal{A}\) as follows
\begin{eqnarray}
\Psi^{i,j,k,l} = \mathcal{A} \left[\phi_L^{i} \otimes \sigma_s^{j} \otimes \zeta_{I}^{k} \otimes \chi_{c}^{l}\right].
\end{eqnarray}
Since the quark components of \(K\bar{K}_1\) and \(KK_1\) are different, the antisymmetrization operators for their wave functions are also different. For \(K\bar{K}_1\), the antisymmetrization operator is \(\mathcal{A} = 1\), while for \(KK_1\), it is \(\mathcal{A} = 1 - (13) - (24) + (13)(24)\).

In addition to the antisymmetrization operator, the main distinction between \(K\bar{K}_1\) and \(KK_1\) also lies in their flavor wave functions \(\zeta_I^k\). The quark components of the \(K\bar{K}_1\) tetraquark system are \(q\bar{s}s\bar{q}\), and it possesses two structures, the molecular structure \(q\bar{s}\)-\(s\bar{q}\), and the diquark structure \(qs\)-\(\bar{s}\bar{q}\). To related with $\eta_1(1855)$ isospin \(I\) of their flavor wave functions is set to 0, and thus the two flavor wave functions are
\begin{eqnarray}
\zeta_0^1 = \frac{1}{\sqrt{2}} \left( u\bar{s}s\bar{u} + d\bar{s}s\bar{d} \right),
\zeta_0^2 = \frac{1}{\sqrt{2}} \left( us\bar{s}\bar{u} + ds\bar{s}\bar{d} \right).
\end{eqnarray}
Similarly, the quark components of the \(KK_1\) tetraquark system are \(q\bar{s}q\bar{s}\), and it also has two structures, the molecular structure \(q\bar{s}\)-\(q\bar{s}\), and the diquark structure \(qq\)-\(\bar{s}\bar{s}\). The flavor wave functions are as follows
\begin{eqnarray}
\zeta_0^3 = \frac{1}{\sqrt{2}} \left( u\bar{s}d\bar{s} - d\bar{s}u\bar{s} \right),
\zeta_0^4 = \frac{1}{\sqrt{2}} \left( ud\bar{s}\bar{s} - du\bar{s}\bar{s} \right).
\end{eqnarray}

For the spins of the quark and antiquark, they are indistinguishable regardless of whether it is a diquark structure or a molecular structure, thus, the spin wave functions of \(K\bar{K}_1\) and \(KK_1\) are the same. For the sub-clusters, the spin wave functions are shown below
\begin{align*}
&\chi_{11}^{\sigma}=\alpha\alpha,~~
\chi_{10}^{\sigma}=\frac{1}{\sqrt{2}}(\alpha\beta+\beta\alpha),~~
\chi_{1-1}^{\sigma}=\beta\beta,\nonumber \\
&\chi_{00}^{\sigma}=\frac{1}{\sqrt{2}}(\alpha\beta-\beta\alpha),
\end{align*}
where \(\alpha\) and \(\beta\) represent the third component of quark spin, taking values of \(\frac{1}{2}\) and \(-\frac{1}{2}\), respectively. By coupling the spin wave functions of the two sub-clusters with Clebsch-Gordan coefficients, the total spin wave function can be written as
\begin{align*}
\sigma_{0}^{1}&=\chi_{00}^{\sigma}\chi_{00}^{\sigma},
\sigma_{0}^{2}=\sqrt{\frac{1}{3}}(\chi_{11}^{\sigma}
  \chi_{1-1}^{\sigma}-\chi_{10}^{\sigma}\chi_{10}^{\sigma}+\chi_{1-1}^{\sigma}\chi_{11}^{\sigma}),\\
\sigma_{1}^{3}&=\chi_{00}^{\sigma}\chi_{11}^{\sigma},
\sigma_{1}^{4}=\chi_{11}^{\sigma}\chi_{00}^{\sigma},
\sigma_{1}^{5}=\frac{1}{\sqrt{2}}(\chi_{11}^{\sigma}\chi_{10}^{\sigma}-\chi_{10}^{\sigma}\chi_{11}^{\sigma}),\\
\sigma_{2}^{6}&=\chi_{11}^{\sigma}\chi_{11}^{\sigma}.\\
\end{align*}

For the color wave functions of \(K\bar{K}_1\) and \(KK_1\), they need to satisfy the color-neutral principle. Therefore, for their molecular state structures, there are two possible configurations, i.e., \(1 \otimes 1 \to 1\), and \(8 \otimes 8 \to 1\). For the diquark-antidiquark structure, the two possible configurations are \(\bar{3} \otimes 3 \to 1\), and \(6 \otimes \bar{6} \to 1\).
\begin{eqnarray*}
\chi_c^1 &= & \sqrt{\frac{1}{9}}(\bar{r}r\bar{r}r+\bar{r}r\bar{g}g+\bar{r}r\bar{b}b+\bar{g}g\bar{r}r+\bar{g}g\bar{g}g \nonumber\\
 & & +\bar{g}g\bar{b}b+\bar{b}b\bar{r}r+\bar{b}b\bar{g}g+\bar{b}b\bar{b}b), \nonumber \\
\chi_c^2 & = & \sqrt{\frac{1}{72}}(3\bar{b}r\bar{r}b+3\bar{g}r\bar{r}g+3\bar{b}g\bar{g}b+3\bar{g}b\bar{b}g+3\bar{r}g\bar{g}r \nonumber \\
& & +3\bar{r}b\bar{b}r+2\bar{r}r\bar{r}r+2\bar{g}g\bar{g}g+2\bar{b}b\bar{b}b-\bar{r}r\bar{g}g \nonumber \\
& & -\bar{g}g\bar{r}r-\bar{b}b\bar{g}g-\bar{b}b\bar{r}r-\bar{g}g\bar{b}b-\bar{r}r\bar{b}b), \nonumber \\
\chi_c^3&= &
 \sqrt{\frac{1}{12}}(rg\bar{r}\bar{g}-rg\bar{g}\bar{r}+gr\bar{g}\bar{r}-gr\bar{r}\bar{g}+rb\bar{r}\bar{b} \nonumber \\
 & & -rb\bar{b}\bar{r}+br\bar{b}\bar{r}-br\bar{r}\bar{b}+gb\bar{g}\bar{b}-gb\bar{b}\bar{g} \nonumber \\
 & & +bg\bar{b}\bar{g}-bg\bar{g}\bar{b}), \nonumber \\
\chi_c^4 &= & \sqrt{\frac{1}{24}}(2rr\bar{r}\bar{r}+2gg\bar{g}\bar{g}+2bb\bar{b}\bar{b}
    +rg\bar{r}\bar{g}+rg\bar{g}\bar{r} \nonumber \\
& & +gr\bar{g}\bar{r}+gr\bar{r}\bar{g}+rb\bar{r}\bar{b}+rb\bar{b}\bar{r}+br\bar{b}\bar{r} \nonumber \\
& & +br\bar{r}\bar{b}+gb\bar{g}\bar{b}+gb\bar{b}\bar{g}+bg\bar{b}\bar{g}+bg\bar{g}\bar{b}).
\end{eqnarray*}

Finally, for the spatial wave function of \(K\bar{K}_1\) and \(KK_1\), it consists of the motion of two sub-clusters and the relative motion between them. Since we consider two spatial structures, i.e., the diquark structure and the molecular structure, the difference in their spatial wave functions lies in the encoding of the particles. For example, in the molecular structure, the encoding of the four-quark spatial wave function is \(q_1 \bar{s}_2\) - \(s_3 \bar{q}_4\), while in the corresponding diquark structure, the encoding of the four-quark spatial wave function is \(q_1 s_3\) - \(\bar{s}_2 \bar{q}_4\). Here, we set the orbital angular momentum of the second sub-cluster to 1 and that of the first sub-cluster to 0. First, we couple them to obtain the orbital quantum number \(l_{12} = 1\), then we couple this spatial wave function with the relative motion wave function to get the total orbital wave function \(\phi_{L=1}^{i}\). The expression is as
\begin{eqnarray}\label{spatialwavefunctions}
\phi_{L}^1 &=& \left[ [\psi_{l_1=0}({\bf r}_{12}) \psi_{l_2=1}({\bf r}_{34})]_{l_{12}=1} \psi_{L_r}({\bf r}_{1234}) \right]_{L=1}, \nonumber \\
\phi_{L}^2 &=& \left[ [\psi_{l_1=0}({\bf r}_{13}) \psi_{l_2=1}({\bf r}_{24})]_{l_{12}=1} \psi_{L_r}({\bf r}_{1324}) \right]_{L=1}, \nonumber \\
\phi_{L}^3 &=& \left[ [\psi_{l_1=1}({\bf r}_{13}) \psi_{l_2=0}({\bf r}_{24})]_{l_{12}=1} \psi_{L_r}({\bf r}_{1324}) \right]_{L=1}.
\end{eqnarray}

For each relative motion wave function, we use the Gaussian expansion method (GEM) for the description, and its form is
\begin{align}
\psi(\mathbf{r}) = \sum_{n=1}^{n_{\rm max}} c_n N_{nl} r^{l} e^{-\nu_n r^2} Y_{lm}(\hat{\mathbf{r}}),
\end{align}
where $N_{nl}$ are normalization constants,
\begin{align}
N_{nl}=\left[\frac{2^{l+2}(2\nu_{n})^{l+\frac{3}{2}}}{\sqrt{\pi}(2l+1)}
\right]^\frac{1}{2}.
\end{align}
The coefficients \(c_n\) are variational parameters, which are determined dynamically. The Gaussian size parameters are chosen according to thegeometric progression
\begin{equation}\label{gaussiansize}
\nu_n = \frac{1}{r_n^2}, \quad r_n = r_1 a^{n-1}, \quad a = \left( \frac{r_{n_{\rm max}}}{r_1} \right)^{\frac{1}{n_{\rm max}-1}}.
\end{equation}

\begin{table*}[tp]
  \centering
  \fontsize{9}{8}\selectfont
  \makebox[\textwidth][c]{
   \begin{threeparttable}
   \caption{\label{Boundstate1} Results of the bound state calculations in the $K\bar{K}_1$ and $K K_1$ system.(unit: MeV)}
    \begin{tabular}{ccccccccc}
\hline\hline
Channel                           &~~~ $\Psi^{i,j,k,l}$~~~  & E     & ~~~~~Percent~~~~~ & Channel                           &~~~ $\Psi^{i,j,k,l}$ ~~~ & E     & ~~~~~Percent~~~~~ \\
 \hline
 $ K\bar{K_1}$                               & $\Psi^{1,3,1,1}$      & $1896$  &$46.6\%$&$ KK_1$                          & $\Psi^{1,3,3,1}$              & $1896$ &$34.5\%$\\
 $ [K\bar{K_1}]_8$                           & $\Psi^{1,3,1,2}$      & $2307$  & $0.0\%$&$ [KK_1]_8$                      & $\Psi^{1,3,3,2}$              & $2305$ & $0.2\%$\\
 $K\bar{K_1'}$                               & $\Psi^{1,1,1,1}$      & $1879$  &$53.4\%$&$KK_1'$                          & $\Psi^{1,1,3,1}$              & $1879$ &$45.5\%$\\
 $[K\bar{K_1'}]_8$                           & $\Psi^{1,1,1,2}$      & $2319$  & $0.0\%$&$[KK_1']_8$                      & $\Psi^{1,1,3,2}$              & $2259$ & $1.7\%$\\
 $K^*\bar{K_0}$                              & $\Psi^{1,5,1,1}$      & $2230$  & $0.0\%$&$K^*K_0$                         & $\Psi^{1,5,3,1}$              & $2230$ & $0.2\%$\\
 $[K^*\bar{K_0}]_8$                          & $\Psi^{1,5,1,2}$      & $2312$  & $0.0\%$&$[K^*K_0]_8$                     & $\Psi^{1,5,3,2}$              & $2309$ & $0.3\%$\\
 $K^*\bar{K_1}$                              & $\Psi^{1,5,1,1}$      & $2320$  & $0.0\%$&$K^*K_1$                         & $\Psi^{1,5,3,1}$              & $2317$ & $0.2\%$\\
 $[K^*\bar{K_1}]_8$                          & $\Psi^{1,5,1,2}$      & $2271$  & $0.0\%$&$[K^*K_1]_8$                     & $\Psi^{1,5,3,2}$              & $2265$ & $2.0\%$\\
 $K^*\bar{K_1'}$                             & $\Psi^{1,4,1,1}$      & $2302$  & $0.0\%$&$K^*K_1'$                        & $\Psi^{1,4,3,1}$              & $2301$ & $0.1\%$\\
 $[K^*\bar{K_1'}]_8$                         & $\Psi^{1,4,1,2}$      & $2310$  & $0.0\%$&$[K^*K_1']_8$                    & $\Psi^{1,4,3,2}$              & $2321$ & $0.7\%$\\
$^0[qs]_3^0$-$^0[\bar{s}\bar{q}]_{\bar{3}}^1$ & $\Psi^{2,1,2,3}$     & $2335$  & $0.0\%$&$^0[qq]_3^0$-$^0[\bar{s}\bar{s}]_{\bar{3}}^1$ & $\Psi^{2,1,4,3}$ & $1978$ & $9.8\%$\\
$^0[qs]_6^0$-$^0[\bar{s}\bar{q}]_{\bar{6}}^1$ & $\Psi^{2,1,2,4}$     & $2307$  & $0.0\%$&$^1[qq]_6^0$-$^1[\bar{s}\bar{s}]_{\bar{6}}^1$ & $\Psi^{2,5,4,4}$ & $2242$ & $0.4\%$\\
$^0[qs]_3^0$-$^1[\bar{s}\bar{q}]_{\bar{3}}^1$ & $\Psi^{2,3,2,3}$     & $2342$  & $0.0\%$&$^1[qq]_3^1$-$^1[\bar{s}\bar{s}]_{\bar{3}}^0$ & $\Psi^{3,5,4,3}$ & $2398$ & $1.0\%$\\
$^0[qs]_6^0$-$^1[\bar{s}\bar{q}]_{\bar{6}}^1$ & $\Psi^{2,3,2,4}$     & $2288$  & $0.0\%$&$^0[qq]_6^1$-$^0[\bar{s}\bar{s}]_{\bar{6}}^0$ & $\Psi^{3,1,4,4}$ & $2341$ & $3.4\%$\\
$^1[qs]_3^0$-$^0[\bar{s}\bar{q}]_{\bar{3}}^1$ & $\Psi^{2,4,2,3}$     & $2447$  & $0.0\%$&\\
$^1[qs]_6^0$-$^0[\bar{s}\bar{q}]_{\bar{6}}^1$ & $\Psi^{2,4,2,4}$     & $2296$  & $0.0\%$&\\
$^1[qs]_3^0$-$^1[\bar{s}\bar{q}]_{\bar{3}}^1$ & $\Psi^{2,5,2,3}$     & $2445$  & $0.0\%$&\\
$^1[qs]_6^0$-$^1[\bar{s}\bar{q}]_{\bar{6}}^1$ & $\Psi^{2,5,2,4}$     & $2244$  & $0.0\%$&\\
 \multicolumn{3}{c}{ complete coupled-channels:}  & $1754$   &\multicolumn{3}{c}{ complete coupled-channels:}  & $1735$\\
\multicolumn{8}{c}{ Threshold($ K$+$K_1(\bar{K}_1)$): $1752$}   \\
\hline\hline
    \end{tabular}
   \end{threeparttable}}
  \end{table*}
\section{Result} \label{result}

In this work, we primarily perform the bound state calculation for the four-quark systems of \(K\bar{K}_1\) (\(q\bar{s}\)-\(s\bar{q}\)) and \(KK_1\) (\(q\bar{s}\)-\(q\bar{s}\)). We will then analyze the contributions from each part of the potential energy for these bound states, as well as the quark distance. Finally, we compare the differences between these two systems.

\subsection{Bound-state calculation of $K\bar{K}_1$ system}

In the \(K\bar{K}_1\) tetraquark system, considering that \(J^{P} = 1^{-}\) and the inter-cluster orbital wave function is S-wave, there are three possible coupling scenarios for the two sub-clusters (\(J_1\) and \(J_2\)), \(J = J_1 \otimes J_2 = 0 \otimes 1\), \(1 \otimes 0\), or \(1 \otimes 1\). Therefore, the possible physical channels are \(K \bar{K}_1\), \(K^* \bar{K}_0\), \(K^* \bar{K}_1\), \(K\bar{K}_1'\), \(K^*\bar{K}_1'\), along with the corresponding 5 color octet states and 8 diquark structures (where we use \(^\text{spin}[qq]^L_{\text{color}} \)- \( ^\text{spin}[\bar{q}\bar{q}]^L_{\text{color}}\) to represent the diquark structures). These are all listed in Table~\ref{Boundstate1}. Our calculations show that the two lowest-energy channels are \(K\bar{K}_1\) and \(K\bar{K}_1'\), with energies slightly below 1.9 GeV. The energies of other color-singlet states are concentrated around 2.3 GeV and correspond to scattering states. Similarly, for the color structures, the energies of the color octet and diquark structures are concentrated in the range of 2.2-2.4 GeV, which is significantly higher than the lowest threshold of the \(K\bar{K}_1\) tetraquark system. Due to this, their channel coupling effects are very small, and the final complete coupled-channel calculation shows that the \(K\bar{K}_1\) tetraquark system does not form a bound state. Therefore, if we do not considering the system mixing between (\(q\bar{s}\)-\(s\bar{q}\)) and (\(q\bar{q}\)-\(q\bar{q}\)) as Refs.~\cite{Yan:2023vbh}, our current calculations do not support the \( \eta_1(1855) \) as a \(K\bar{K}_1\) molecular state.

\subsection{Bound-state calculation of $K K_1$ system}

Similar to the \(K\bar{K}_1\) tetraquark system, the \(KK_1\) tetraquark system also has five color-singlet channels as \(K K_1\), \(K^* K_0\), \(K^* K_1\), \(K K_1'\), and \(K^* K_1'\), as well as the corresponding 5 color octet states. The difference from the \(K\bar{K}_1\) tetraquark system is that, due to symmetry constraints, the \(KK_1\) tetraquark system has only four diquark structures as \(\Psi^{2,1,4,3}\), \(\Psi^{2,5,4,4}\), \(\Psi^{3,5,4,3}\), and \(\Psi^{3,1,4,4}\). These are all listed in Table~\ref{Boundstate1}. Our single-channel calculation results show that the lowest energy channels are still \(K K_1\) and \(K K_1'\), with the color octet states concentrated between 2.2 GeV and 2.3 GeV. In the diquark-antidiquark structures, the energy of \(\Psi^{2,1,4,3}\) \(( ^0[qq]_3^0 - ^0[\bar{s}\bar{s}]_{\bar{3}}^1 )\) is relatively low at 1.98 GeV, while the energies of the others are between 2.2 GeV and 2.3 GeV. This occurs because in the diquark structure \(\Psi^{2,1,4,3}\), the \(qq\) cluster of \(( ^0[qq]_3^0 - ^0[\bar{s}\bar{s}]_{\bar{3}}^1 )\) forms a good-diquark structure, meaning the quantum numbers of \(qq\) are \(00^{+}\) and the color wave function is antisymmetric. Due to the presence of such a low-energy diquark channel close to the lowest threshold, the final complete coupled-channel calculation shows that we obtain a bound state with an energy of 1735 MeV. The main components of this bound state are \(K K_1\) (about $34.5\%$), \(K K_1'\) (about $45.5\%$), and \(^0[qq]_3^0\)-\(^0[\bar{s}\bar{s}]_{\bar{3}}^1\) (about $9.8\%$).

\subsection{Comparison between the \(K K_1\) system and the \(K \bar{K}_1\) system}
\begin{figure}[htp]
  \setlength {\abovecaptionskip} {0.3cm}
  \centering
  \resizebox{0.30\textwidth}{!}{\includegraphics[width=2.0cm,height=1.5cm]{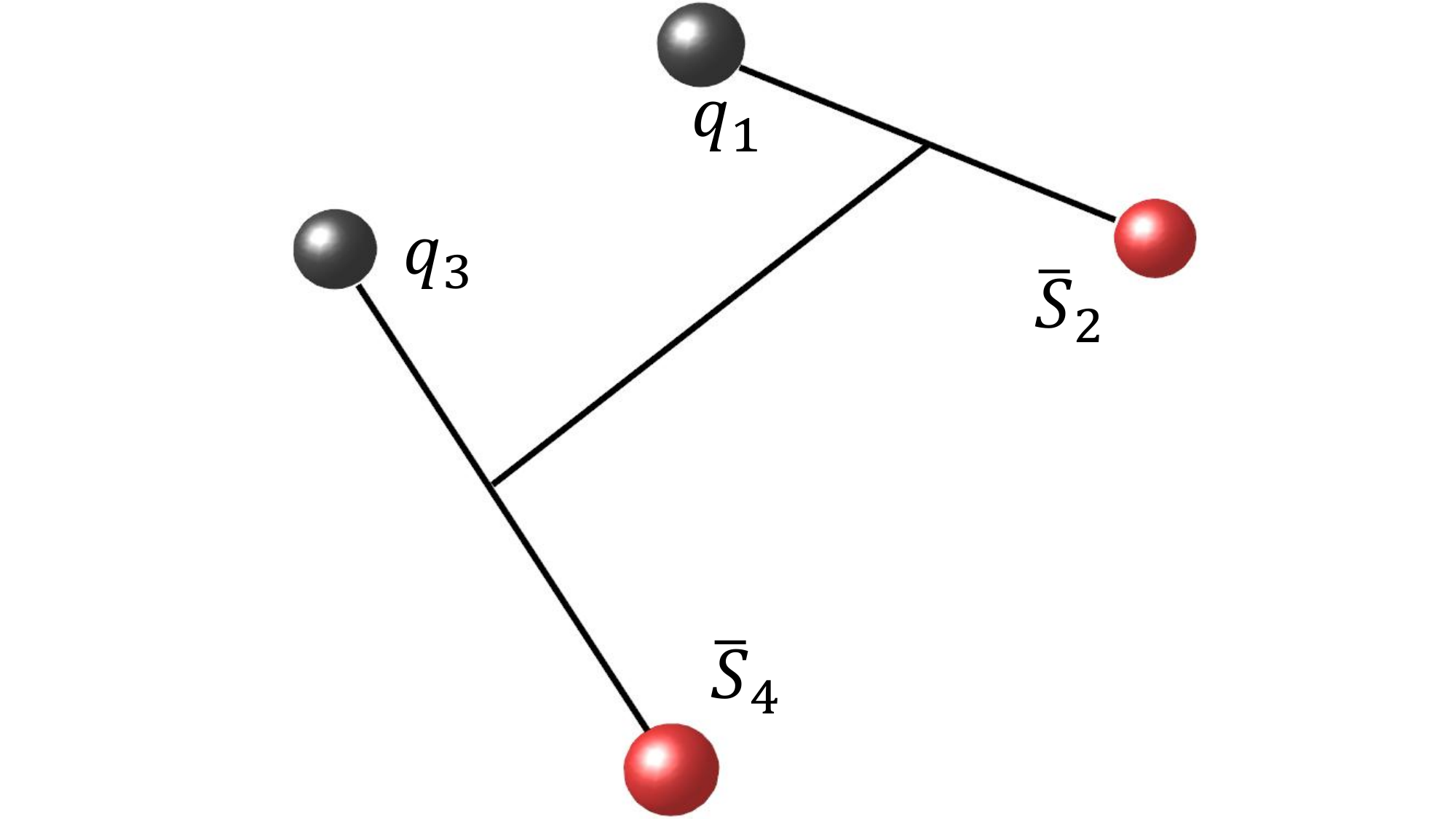}}
  \caption{The spatial configuration of $KK_1$.}
\label{sptial}
\end{figure}

To compare the differences between the \(K K_1\) system and the \(K \bar{K}_1\) system, we first calculate the contribution of the potential energy for each part of the lowest energy of these two systems (by subtracting the potential energy contributions of the threshold state from the results of the lowest energy state) and the root-mean-square (rms) distances, which are listed in Table~\ref{rms}. Since \(K \bar{K}_1\) is a scattering state, the contributions from each part of the potential energy are relatively small. Although the \(\sigma\)-meson still provides about 5 MeV of binding energy, its kinetic energy is repulsive, reaching 7.7 MeV, and thus no bound state is formed. On the other hand, as seen in Table~\ref{rms}, the rms distance between quarks in the bound \(K K_1\) state is smaller, and its kinetic energy is relatively large, reaching 104.7 MeV. However, due to the smaller quark distances, the meson exchange contribution to the binding energy is relatively large, especially the \(\pi\)-meson exchange, which reaches 91.1 MeV.  Because \(q_1\) and \(q_3\), as well as \(\bar{s}_2\) and \(\bar{s}_4\), are identical particles. Therefore, we only list the three rms distances \(r_{q\bar{s}}\), \(r_{q q}\), and \(r_{\bar{s}\bar{s}}\) in Table~\ref{rms}. However, these values do not represent the true distances between particles but rather an averaged effect. Using a triangular approximation (\(r_{q_1 \bar{s}_2}^{2} = \frac{r_{re}^{2} + r_{q_1 q_3}^{2}}{2}\)), where \(r_{re}\) is the real rms of \(r_{q_1 \bar{s}_2}\), we can obtain a cluster of one meson (\(q_1\)-\(\bar{s}_2\)) with a distance of about 0.64 fm. Similarly, the actual distance between \(q_3\) and \(\bar{s}_4\) is 1.12 fm. Based on this actual distance, we plot the spatial configuration of the bound \(K K_1\) state, shown in FIG.~\ref{sptial}. We can see that since \(K_1\) is an excited state, \(q_3\) and \(\bar{s}_4\) are relatively farther apart, while \(q_1\) and \(q_3\) form a good diquark structure, resulting in a very small distance between them.

\begin{table*}[htb]
\caption{\label{rms} The root-mean-square distances (unit: fm) and  contributions of all potentials to the binding energy (unit: MeV) in $K\bar{K}_1$ and $KK_1$ four-quark systems.}
\begin{tabular}{cccccccccccccccccccccccccccc}\hline\hline
&knetic&confinement&coulomb&color-magnetic&~~~$\pi$~~~&~~~$\eta$~~~&~~~ $\sigma$ ~~~&&$r_{q\bar{s}}$ &$r_{qs}$ &$r_{q\bar{q}}$ &$r_{s\bar{s}}$&$r_{\bar{s}\bar{q}}$ &$r_{s\bar{q}}$ \\
$K\bar{K}_1$ &$7.7$&$-0.7$&$-0.1$&$-0.2$&$-0.1$&$-0.1$&$-5.0$& &$0.5$    &$6.8$    &$6.8$    &$6.8$    &$6.8$    &$1.15$  \\
&&&&&&&&&$r_{q\bar{s}}$ &$r_{qq}$ &$r_{\bar{s}\bar{s}}$  \\
$KK_1$       &$104.7$&$-30.2$&$5.3$&$6.9$&$-91.1$&$27.2$&$-56.4$& &$0.9$    &$0.6$    &$1.1$    \\
\hline\hline
\end{tabular}
\end{table*}

\section{Summary} \label{Summary}
Within the framework of the chiral quark model, we systematically studied two four-quark systems, the \(K K_1\) system and the \(K \bar{K}_1\) system. We considered not only the molecular state structures, but also the diquark structures and their mixing effects.

Our calculation results show that in the \(K \bar{K}_1\) system, all the color-singlet channels are scattering states, and the energies of the color structures, including the color octet and diquark structures, are relatively high and far from the system's lowest threshold state. Therefore, despite the involvement of up to 18 physical channels in the coupled-channel calculation, no bound state was obtained. In contrast, due to symmetry constraints, the \(K K_1\) system has only four diquark channels, and one of the diquark structures forms a good diquark structure, which significantly lowers its energy compared to other color structures. This good diquark structure couples strongly with the two lowest-energy channels, \(K K_1\) and \(K K_1'\), leading to the formation of a bound state. The rms distance calculations show that this bound state is a compact four-quark structure, and the distance between the internal \(qq\) quark pairs is very small. Due to the compactness of the four-quark structure, the binding energy of this state is predominantly contributed by meson exchanges, especially $\pi$-exchange and $\sigma$ exchange.

As all, our current calculations do not support the interpretation of \(\eta_1(1855)\) as a \(K \bar{K}_1\) molecule, and it is perhaps similar to \(X(3872)\) in the unquenched quark model \cite{Tan:2019qwe}, requiring the mixing effect of a \(c\bar{c}\) component. In the \(K \bar{K}_1\) tetraquark system, similar mixing effects may be needed, such as the contribution from the D-wave, or a system mixing between \((q\bar{s})-(s\bar{q})\) and \((q\bar{q})-(q\bar{q})\) as done in Ref.~\cite{Yan:2023vbh}, or the mixing effect between \((q\bar{s})-(s\bar{q})\) and \(s\bar{s}g\). However, our results strongly support the existence of a new bound state in the \(K K_1\) system, and we suggest future experiments to search for it.

\acknowledgments{This work is supported partly by the National Science Foundation of China under Contract No. 12205249 and No. 12305087. Y. T. is supported by the Funding for School-Level Research Projects of Yancheng Institute of Technology under Grant No. xjr2022039. Q. H. is supported by the Start-up Funds of Nanjing Normal University under Grant No. 184080H201B20. X.-H. H. is supported by The Programme of Natural Science Foundation of the Jiangsu Higher Education Institutions under No. 1020242167.}

\end{document}